\documentclass{aa} 
\usepackage{graphicx}
\usepackage{txfonts}
\begin{document}
   \title{A multisite photometric campaign on the Pre-Main-Sequence 
         $\delta$ Scuti pulsator IP Per}

\author{
V. Ripepi\inst{1}\and
S. Bernabei\inst{2,3} \and
M. Marconi\inst{1}\and
F. Palla\inst{4}\and  
A. Arellano Ferro\inst{5}\and
A. Bonanno\inst{6}\and
P. Ferrara\inst{1}\and
A. Frasca\inst{6}\and
X.J. Jiang\inst{7}\and 
S.-L. Kim\inst{8}\and  
S. Marinoni\inst{2}\and
G. Mignemi\inst{9}\and
M.J.P.F.G. Monteiro\inst{10}\and
T.D. Oswalt\inst{11}\and
P. Reegen\inst{12}\and
R. Janulis\inst{13}\and
E. Rodriguez\inst{14}\and
A. Rolland\inst{14}\and
A. Ruoppo\inst{1,15} \and
L. Terranegra\inst{1}  \and
K. Zwintz\inst{12}
          }

   \offprints{V. Ripepi}

\institute{
INAF-Osservatorio Astronomico di Capodimonte, 
Via Moiariello 16, 80131, Napoli, Italy 
\and
INAF-Osservatorio Astronomico di Bologna, Via Ranzani 1, 
40127 Bologna, Italy 
\and
Departimento de Astrof\'{\i}sica, Universidad de La Laguna, Avda. 
Astrofisico F. S\'anchez sn, 30071 La Laguna, Spain 
\and 
INAF-Osservatorio Astrofisico di Arcetri, Largo E. Fermi, 5, I-50125
Firenze, Italy 
\and
Instituto de Astronom{\'\i}a, UNAM, Apdo. Postal 70-264, M\'exico D.F.,
CP 04510, M\'exico 
\and
INAF-Osservatorio Astrofisico di Catania, Citt\`a Universitaria, 
95125 Catania, Italy 
\and
National Astronomical Observatories, Chinese Academy of Sciences, 
Beijing, 100012, China 
\and
Korea Astronomy and Space Science Institute, Daejeon 305-348, Korea 
\and
Dipartimento di Fisica e Astronomia dell'Universit\'a, Sezione Astrofisica, 
Citt\'a Universitaria, 95123 Catania, Italy.
\and
DMA-Faculdade de Ci\^encias and Centro de Astrof\'{\i}sica da 
Universidade do Porto, Rua das Estrelas, 4150-762 Porto, Portugal 
\and
Florida Institute Technology, 150 W Univ. Blvd., Melbourne, FL 32901-6988, USA
\and
Institute for Astronomy (IfA), University of Vienna,
T\"urkenschanzstrasse 17, A-1180 Vienna, Austria
\and
Institute of Theoretical Physics \& Astronomy, Vilnius University,
Gostauto 12 Vilnius, Lithuania
\and
Instituto de Astrof\'isica de Andaluc\'ia, CSIC, Apdo. 3004, 18080 Granada, Spain 
\and
Dipartimento di Scienze Fisiche, Universit\`a Federico II, Complesso 
Monte S. Angelo, 80126, Napoli, Italy 
}

   \date{}

   \abstract{We present the results of a photometric multisite
campaign on the $\delta$ Scuti Pre-Main-Sequence star IP Per.  Nine
telescopes have been involved in the observations, with a total of
about 190 hours of observations over 38 nights. Present data confirms
the multiperiodic nature of this star and leads to the identification
of at least nine pulsational frequencies. Comparison with the
predictions of linear non-adiabatic radial pulsation models allowed us
to identify only five of the nine observed frequencies, and to
constrain the position of IP Per in the HR diagram. The latter is in
good agreement with the empirical determination of the stellar
parameters obtained by \cite{miro}.  An initial interpretation of the
observed frequencies using the Aarhus non-radial pulsation code
suggests that three frequencies could be associated with non-radial
($l$=2) modes. \par Finally, we present new evolutionary and pulsation
models at lower metallicity (Z=0.008) to take into account the
possibility that IP Per is metal deficient, as indicated by
\cite{miro}.

   }

   \maketitle
%

\section{Introduction}

Intermediate mass (M $\geq$ 1.5 M$_{\odot}$) Pre-Main-Sequence (PMS)
stars are known as Herbig Ae/Be stars (\cite{herbig}). This class of
stars is characterized by spectral type A or B with emission lines, an
infrared excess due to hot or cool circumstellar dust or both, and
luminosity class III to V (\cite{waters}).  Herbig Ae/Be are also well
known for their photometric and spectroscopic variability on time
scales of minutes to years mainly due to photospheric activity and
interaction with the circumstellar environment (see
e.g. \cite{catala}).  However, the fact that these young stars during
their contraction towards the Main-Sequence (MS) move across the
pulsation instability region of more evolved stars has prompted the
suggestion that at least part of the activity could be due to stellar
pulsation (see \cite{baade}; \cite{kurtz}).

\begin{table}
\caption{Journal of the observations. Note that the data coming from
Sierra Nevada and San Pedro Martir (SPM) have been transformed in the
Johnson system as discussed in section 2.2.}
\label{jou}
\begin{tabular}{cccc}
\hline
\noalign{\smallskip}
HJD-2450000 & HJD-2450000  & Duration &  Filter \\
  start (days)         &   end (days)        &  (hours) &         \\
\noalign{\smallskip}
\hline
& \bf{Loiano CCD  (Italy)}  & & \\
\noalign{\smallskip}
2545.585 &	2545.659&  1.8 & B \\
2547.599 &	2547.681&  2.0 & B\\
2548.547 &	2548.611&  1.5 & B\\
2549.503 &	2549.623&  2.9 & B\\
2628.347 &	2628.627&  6.7 & B\\
2629.286 &	2629.453&  4.0 & B\\
2630.454 &	2630.513&  1.4 & B\\
2654.252 &	2654.332&  1.9 & B\\
\noalign{\smallskip}
 	 &\bf{Fairborn-APT (USA)}	&      &\\
\noalign{\smallskip}
2976.875 &	2976.892&   0.4&  BV\\
2977.851 &	2977.936&   2.0&  BV\\
2978.721 &	2978.933&   5.1&  BV\\
2983.573 &	2983.920&   8.3&  B \\
2984.711 &	2984.917&   4.9&  BV\\
2987.765 &	2987.902&   3.3&  BV\\
\noalign{\smallskip}
	 &\bf{SOAO (Korea)}    &      &\\
\noalign{\smallskip}
2977.170 &	2977.333&   3.9&  BV\\
2980.910 &	2981.063&   3.7&  V\\
2983.007 &	2983.191&   4.4&  BV\\
\noalign{\smallskip}
	 & \bf{OSN (Spain)}	&      &\\
\noalign{\smallskip}
2988.267 &	2988.500&   5.6&  uvby\\
3047.364 &	3047.472&   2.6&  uvby\\
3048.299 &	3048.459&   3.8&  uvby\\
\noalign{\smallskip}
         &\bf{Loiano TTCP (Italy)}	&      &\\
\noalign{\smallskip}
2946.330 &	2946.703&   9.0&  BV\\  
2947.287 &	2947.441&   3.7&  BV\\
2973.393 &	2973.636&   5.8&  BV\\
\noalign{\smallskip}
	 &\bf{SARA (USA)} &      &\\
\noalign{\smallskip}
2970.631 &	2970.756&   3.0&  B\\
\noalign{\smallskip}
	 &\bf{SPM (Mexico)}    &      &\\
\noalign{\smallskip}
2970.494 &	2970.818&  7.8 & uvby\\
2972.495 &	2972.841&  8.3 & uvby\\
2973.493 &	2973.775&  6.8 & uvby\\
2974.487 &	2974.585&  2.4 & uvby\\
2977.485 &	2977.847&  8.7 & uvby\\
\noalign{\smallskip}
         & \bf{BAO (China)}	&      &\\
\noalign{\smallskip}
2972.974 &	2973.383&  9.8 & V\\
2974.032 &	2974.146&  2.7 & V\\
2974.939 &	2975.382& 10.6 & V\\
2976.940 &	2977.078&  3.3 & V\\
2984.196 &	2984.292&  2.3 & V\\
\noalign{\smallskip}
	 &\bf{Teide-OGS (Spain)}&     &\\
\noalign{\smallskip}
2977.506 &	2977.603&  2.3 & V\\
\noalign{\smallskip}
 & \bf{Serra la Nave (Italy)}	  	&      &\\
\noalign{\smallskip}
2966.462 &	 2966.62&  3.9&  BV\\
2967.337 &	 2967.61&  6.7&  BV\\
2973.348 &	 2973.61&  6.4&  BV\\
\noalign{\smallskip}
\hline
\end{tabular}
\end{table}

The possible presence of pulsators among Herbig Ae/Be stars is
particularly attractive since the precise observables which can be
measured, i.e.  the pulsation frequencies can, in principle, allow us
to test evolutionary models by constraining the internal structure
using asteroseismological techniques.

The existence of pulsating Herbig stars was originally suggested by
\cite{breger1} who discovered two candidates in the young open cluster
NGC 2264. This initial finding was confirmed by subsequent
observations of $\delta$ Scuti-like pulsations in the Herbig Ae stars
HR5999 (\cite{kurtz}) and HD104237 (\cite{donati}).\par

This empirical evidence stimulated the first theoretical investigation
of the PMS instability strip based on non-linear convective
hydrodynamical models (Marconi \& Palla 1998).  As a result, the
topology of the PMS instability strip for the first three radial modes
was identified.  \cite{marconi} also pointed out that the interior
structure of PMS stars entering the instability strip differs
significantly from that of more evolved Main Sequence stars (with the
same mass and temperature), even though the envelopes structures are
similar. This property was subsequently confirmed by \cite{suran} who
made a comparative study of the seismology of a 1.8 $M_{\odot}$ PMS
and post-MS star. \cite{suran} found that the unstable frequency range
is roughly the same for PMS and post-MS stars, but that some
non-radial modes are very sensitive to the deep internal structure of
the star. In particular, it is possible to discriminate between the
PMS and post-MS stage using differences in the oscillation frequency
distribution in the low frequency range ($g$ modes, see also
\cite{templeton}). \par

Up to now new observational programs have been carried out by various
groups. The current number of known or suspected candidates amounts to
about 30 stars (see the updated list at
http://ams.astro.univie.ac.at/pms\_corot.php, and the reviews by
\cite{zwintz}, \cite{marconi2004} and \cite{marconi2004a}).  However,
only a few stars have been studied in detail, so that the overall
properties of this class of variables are still poorly determined.

In this context, our group has started a systematic monitoring program
(see \cite{marconi2001}, \cite{h254}, \cite{v346}, \cite{v351},
\cite{bernabei}) of Herbig Ae stars with spectral types from A to F2-3
with the following aims: 1) to identify the largest number of
pulsating objects in order to observationally determine the boundaries
of the instability strip for PMS $\delta$ Scuti pulsation; 2) to study
in detail through multisite campaigns selected objects showing
multiperiodicity (see \cite{marconi2001}, \cite{h254},\cite{v346},
\cite{v351}, \cite{bernabei}).  The multiperiodic pulsators are
potential candidates for future asteroseismological analysis.

\begin{table}

\caption[]{List of telescopes and instruments involved in the
multisite campaign. SCP means Single Channel Photometer; TCP means
Three Channel Photometer. To give an idea of the relative photometric
precision of each site, the last column shows the white noise in the
Fourier space (amplitude) for each site. All the calculations are in
the V filter except for Loiano (2002) and SARA sites for which we have
only B filter data.}
\label{tab1}
\begin{tabular}{lccc}
\hline
\noalign{\smallskip}
Observatory  & Telescope & Instrument & noise \\
& & & (mmag)\\
\noalign{\smallskip}
\hline
\noalign{\smallskip}
Loiano (Italy)    & 1.5m  &  BFOSC- 2002  & 0.2 \\
Loiano (Italy)    & 1.5m  &  TTCP - 2003  & 0.2 \\
BAO (China)       & 0.85m &  TCP & 0.4 \\
SPM (Mexico) & 1.5m  & $uvby$ Phot. & 0.6\\
SARA (USA)  & 0.9m  & CCD  &  1.5 \\
Teide (Spain)     & 1.0m OGS &  CCD   & 0.6 \\
Fairborn (USA)    & 0.75m T6 & SCP  & 0.8 \\
OSN (Spain) & 0.9m &  $uvby$ Phot.  & 0.6\\
Serra la Nave (Italy) & 0.9m & SCP & 1.3 \\
SOAO (Korea)         & 0.6m & CCD  & 0.7 \\ 
\noalign{\smallskip}
\hline
\end{tabular}

\end{table}

During this project our attention turned to the star IP Per, already
listed as Herbig Ae star by \cite{the} with spectral type A3e, and
studied in detail by \cite{miro} both photometrically and
spectroscopically.  The main properties of this interesting object
are:

\begin{itemize}

\item
IP Per is a typical UX Ori type star showing photometric variations
with an amplitude of $\sim$0.3 mag and with duration of the minima of
10-50 days.

\item
the fundamental stellar parameters are: spectral type A7, $T_{\rm eff}
\sim 8000\pm100$ K, $\log g \sim 4.4 \pm 0.1$, $\xi_t=2.0\pm1.0$ km
s$^{-1}$ and [M/H] $\simeq -0.41\pm0.1$.  Thus, IP Per is a dwarf
metal-poor star.

\item
its radial velocity and proper motion suggest that it most likely
belongs to the Per OB2 association ($D \simeq$ 300 pc, \cite{zeeuw}).
At $D \simeq$ 300 pc and $\log L/L_{\odot}$=1.0, IP Per falls onto the
ZAMS at a mass of 1.8 M$_{\odot}$.

\end{itemize}

In the following, we present the results of a photometric multisite
campaign to study in detail the pulsational properties of IP Per.  In
Section 2. we describe the observations and data reduction techniques;
in Section 3. we discuss the frequency analysis; Section 4 presents a
comparison with theoretical models. A brief discussion concludes the
paper. Preliminary results of our observations have been presented in
\cite{ripmarc} and \cite{ripepiiau}.


\section{Observations and data reduction}

The campaign on IP Per was conducted in two parts: 1) single site
observations during winter 2002/2003 (Loiano telescope); 2) multisite
campaign during winter 2003/2004, involving 9 different
telescopes/instruments, as described in Table~\ref{tab1} and more in
detail in the next section.  A total of about 190 hours of
observations spanning 38 nights have been gathered. A detailed log of
the observations is shown in Table~\ref{jou}.

As for the comparison star, we typically used the star TYC 2359-802-1
(03$^h$ 41$^m$ 15.36$^s$ +32$^{\circ}$ 23$^{'}$ 49.7$^{''}$ (2000)
B$_T$=12.029 mag, V$_T$=11.299 mag).  It has been checked on the basis
of CCD data obtained in Loiano in 2002, see below). HD 278941 (03$^h$
41$^m$ 01.67$^s$ +32$^{\circ}$ 07$^{'}$ 06.1$^{''}$ (2000)
B$_T$=9.957, V$_T$=9.576 mag, A5) was the check star.

\subsection{Details on single site observations}

The exploratory run on IP Per was carried out in 2002/2003 with the
Loiano 1.5m telescope (Italy).  The observations has been gathered in
the B filter by using the BFOSC instrument
(http://www.bo.astro.it/loiano/observe.htm\#manuals) equipped with a
CCD EEV 1300x1340 pixel (R.O.N.=1.73 e-/pixel; GAIN=2.13 e-/ADU).  The
pixel scale was 0.58 arcsec/pixel, for a total field of view of 12.6
$\times$ 13 acrmin$^{2}$.  Sky flats, dark and bias exposures were
taken every night.  All data were reduced using standard IRAF
routines. Aperture photometry was carried out by using the DAOPHOT II
package (\cite{stet87}). \par

The multisite campaign in 2003 was carried out by using a variety of
instruments, as described in some details in the Appendix.

   \begin{figure}
   \centering
   \includegraphics[height=8cm]{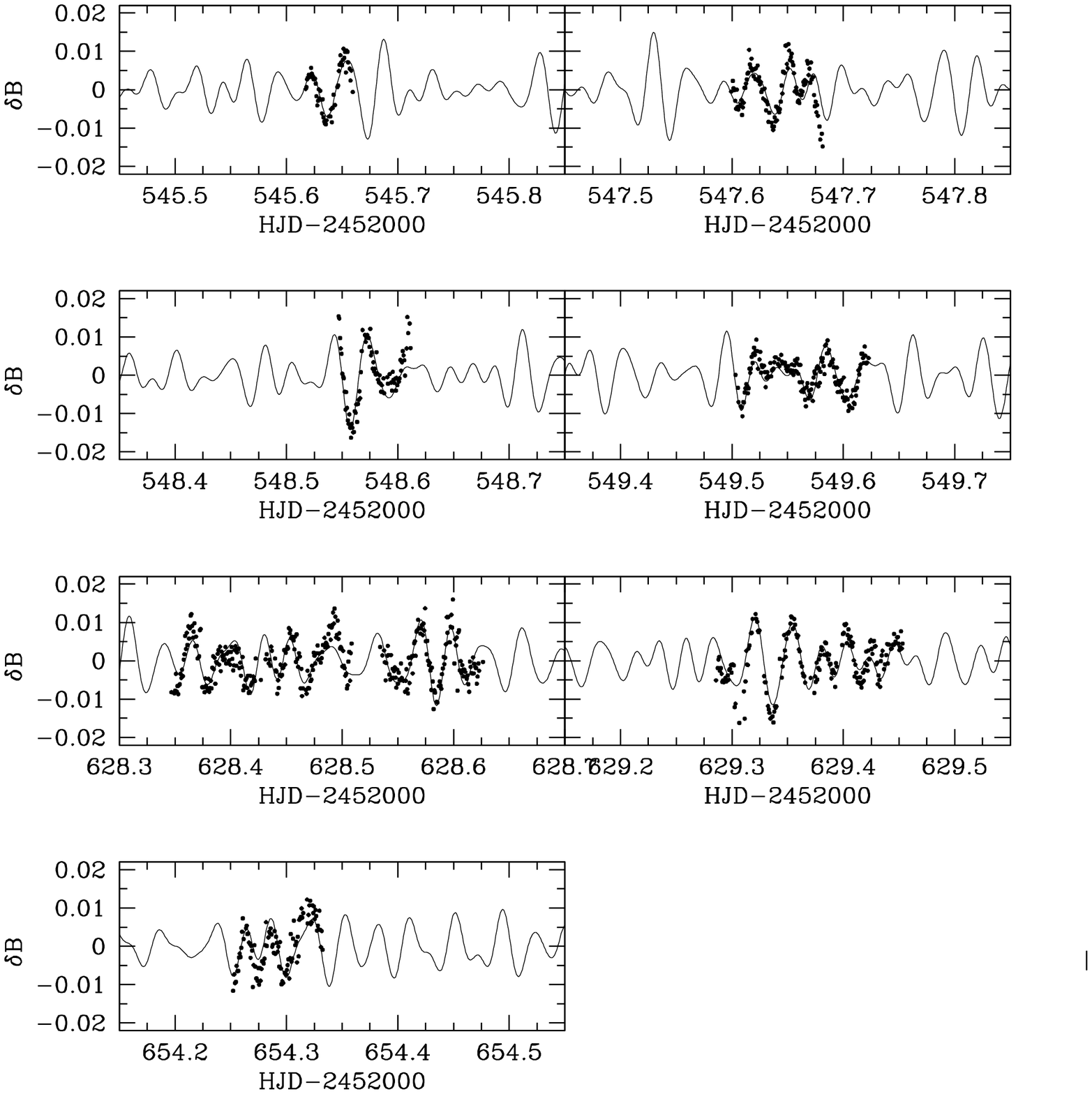}
      \caption{Light curves for the 2002 B dataset (Loiano). Note that
$\delta B=B_{VAR}-B_{COMP}$. The solid line represents the fit to the
data including the six significant frequencies found for this dataset
 (see Sect.3).}
      \label{f1a}
   \end{figure}
%

\subsection{Merging the data from different sites}

In order to use simultaneously all the data gathered during the
campaign we had to convert the data coming from Sierra Nevada and San
Pedro Martir sites from Str\"omgren $uvby$ filters to the Johnson $B$
and $V$ ones.  This was straightforward for the $V$ because $\delta V
\approx \delta y $ (where $\delta V$ means
$V_{variable}-V_{comparison}$), whereas to obtain $\delta B$ we used
the linear approximation by \cite{warren} (equation 5), obtaining
$\delta B \simeq \delta y +1.668 \delta (b-y)$.

   \begin{figure*}
   \centering
\hbox{
   \includegraphics[height=9.2cm]{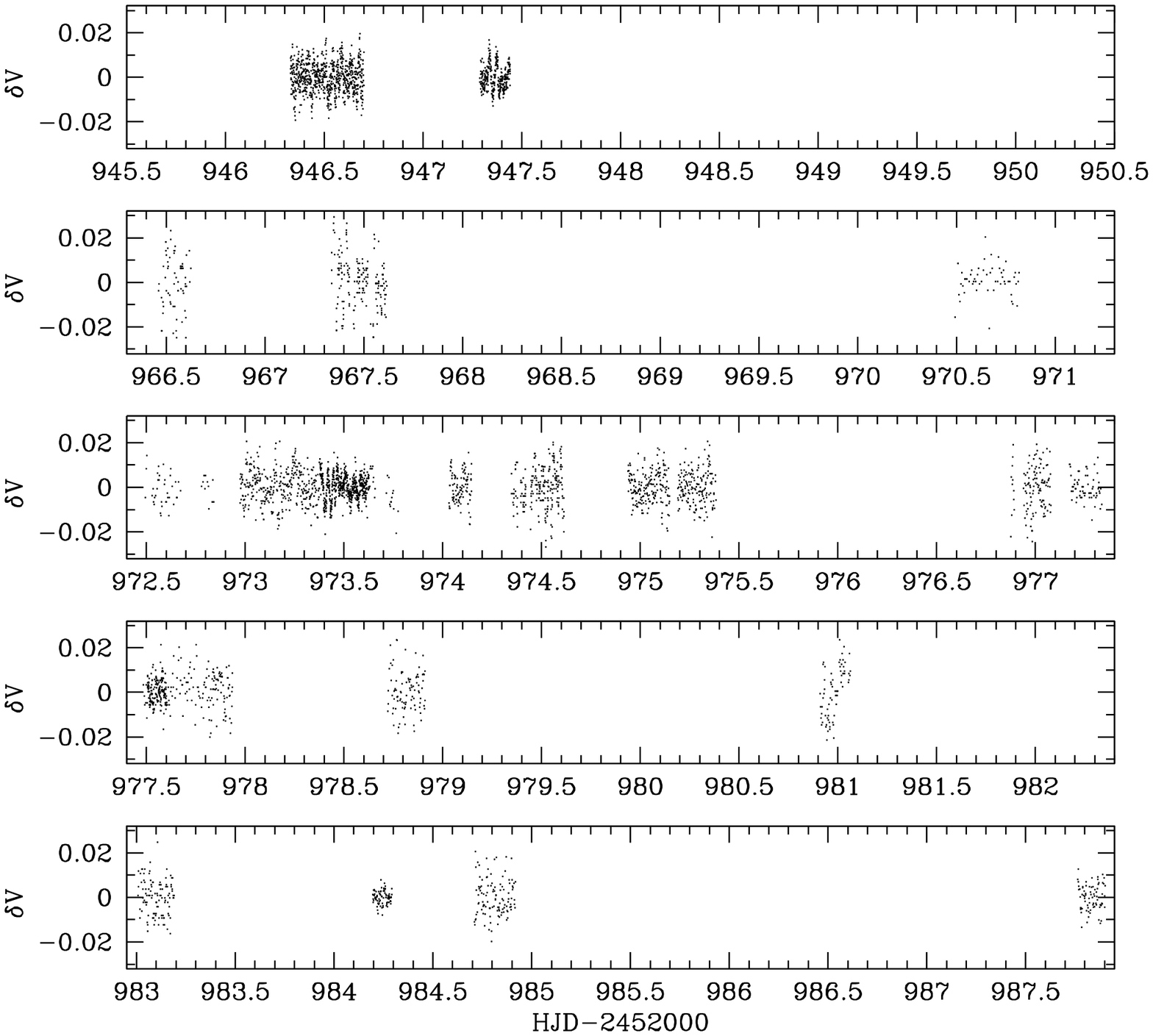}
   \includegraphics[height=9.2cm]{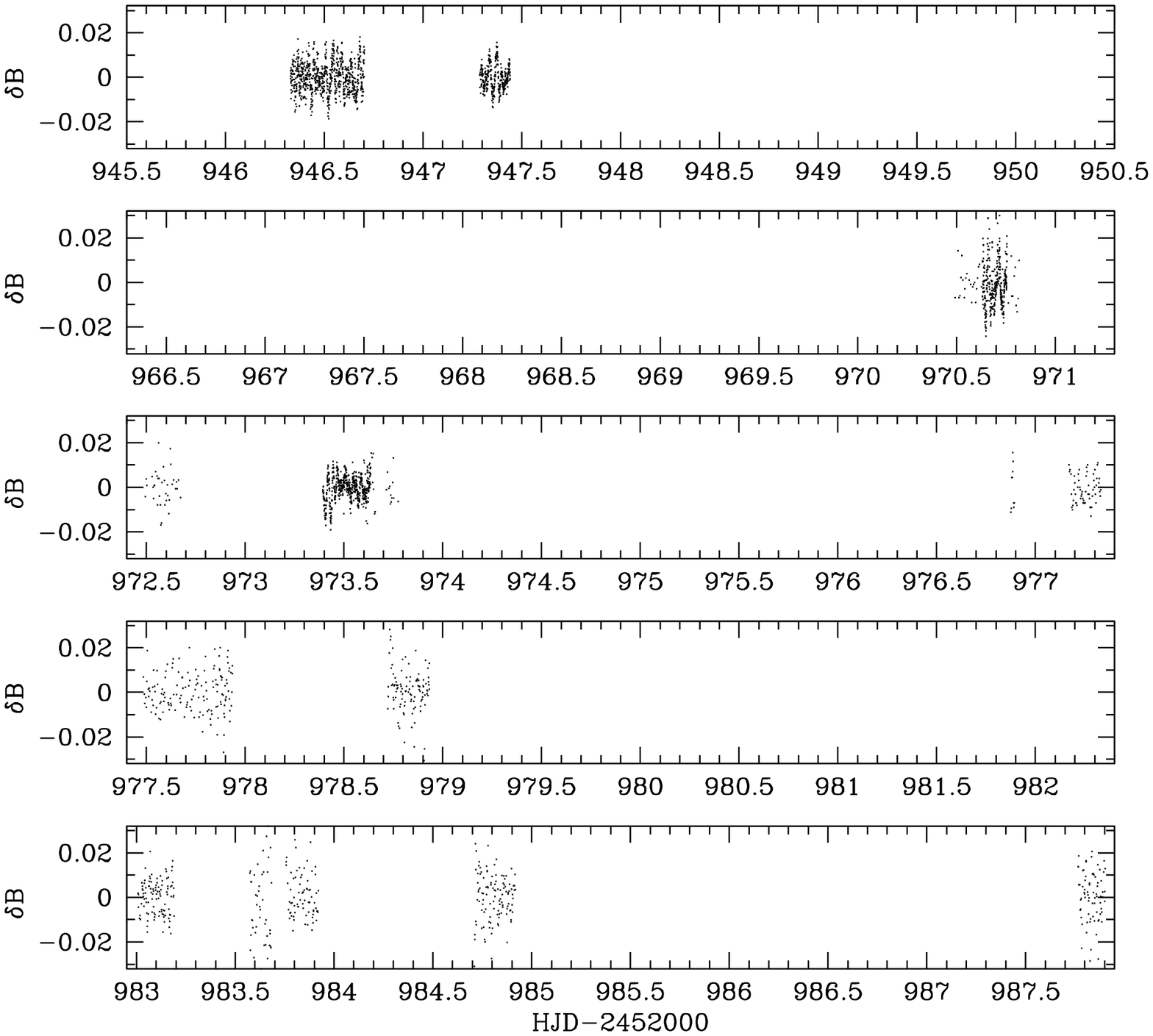}
}
      \caption{Left panels: light curves for the 2003 V dataset. Note that 
$\delta V=V_{VAR}-V_{COMP}$. Right panels: light curves for the 2003 B 
dataset. Note that $\delta B=B_{VAR}-B_{COMP}$}
      \label{f1b}
   \end{figure*}
%

The different datasets from the various telescopes were inspected carefully,
and scattered points due to bad weather (very frequent for data
obtained with single channel photometers data) or other causes have
been removed.

Due to the differences of the various instruments and filters used,
and in order to prevent problems with zero point differences between
different datasets, as well as to prepare the data for Fourier analysis, we
decided to detrend the data to a common average zero value.

In total, during our study of IP Per we obtained three different time series:
\begin{enumerate}
\item
B filter data from Loiano 2002/2003 observations (2002 B dataset here after)
\item
B filter data from multisite campaign 2003/2004 (2003 B dataset here after)
\item
V filter data from multisite campaign 2003/2004 (2003 V dataset here after)
\end{enumerate}
 
The light curve for the 2002 B dataset is shown in Fig.~\ref{f1a}, 
while the left and right panels of Fig.~\ref{f1b} show the light curves for the 
2003 V and B datasets respectively. In order to appreciate 
the quality of the V data taken
in 2003 we report in Fig.~\ref{f2} a zoom of Fig.~\ref{f1b} (left
panel) including only the Loiano TTCP data (V filter), which is
representative of our highest quality photometry.

\subsection{{\em uvby} H$_{\beta}$ measurements for IP Per}

During our {\em uvby} measurements collected at Sierra Nevada
Observatory, two brighter comparison stars C1=HD~22418 and C2=HD~21913
were used for purposes of calibration. In addition, a few H$_{\beta}$
observations were also obtained.  Then, instrumental magnitude
differences were obtained relative to C1. To transform these
instrumental differences into the standard {\em uvby$_{\beta}$}
system, we have followed the procedure described in
\cite{rod97}. Thus, the transformation equations, obtained using a set
of 19 standards stars selected from the list of \cite{craw66} and
\cite{craw70}, were used. \par

Next, the absolute standard {\em uvby$_{\beta}$} indices of IP~Per, C1
and C2 were obtained following the method described in \cite{rod03},
using C1 and C2 as zero-points. The results are listed in
Table~\ref{eloy} together with those listed in the bibliography for
the comparison stars. The error bars in this table represent the standard
deviations of magnitude differences relative to C1. As seen, our
results are in very good agreement with the values found in the
homogeneous catalogue of Olsen (1996, private communication).  Similar
results can be found in the list by \cite{olsen83} and \cite{hauck}.

\begin{table} 
\caption[]{{\em uvby$_{\beta}$} indices obtained for IP~Per and
comparison stars. The pairs below the star names are the number of
points collected for each object in {\em uvby} and $\beta$,
respectively. The numbers below the measures show the errors (in
mmag). The values given by Olsen (1996, private communication) are
listed in the bottom part.}
\label{eloy}
\begin{tabular}{lccccc}
\hline
\noalign{\smallskip}
Star      &     V   &   b-y  &   m$_1$  &   c$_1$  & $\beta$ \\
          &   (mag)  & (mag) &   (mag)  &   (mag)  &  (mag)  \\
\noalign{\smallskip}
\hline
\noalign{\smallskip}
IP Per     &  10.374  & 0.239 &   0.144  &   0.834   &  2.763 \\
(49,5)     &       6  &    6  &      6   &     15    &     9   \\
C1=HD22418 &   6.963  & 0.280 &   0.163  &   0.420  &   2.646  \\
(49,6)     &       4  &     3 &       4  &       9  &       4 \\   
C2=HD21913 &   7.627 & 0.297  &  0.136  &   0.473   &  2.640 \\
(21,5)     &       4 &     3  &      4  &       9   &      5  \\
\noalign{\smallskip}
\hline
\noalign{\smallskip}
C1=HD22418  &  6.962 & 0.281  &  0.167  &  0.419  &   2.645 \\
C2=HD21913  &  7.628 & 0.296  &  0.132 &    0.473 &    2.640 \\
\noalign{\smallskip}
\hline
\end{tabular}

\end{table}

\begin{table*}
\caption{Frequencies, amplitudes and confidence levels for the three
datasets analysed in this paper. The error bars on the frequencies are
$\pm$0.11 c/d, $\pm$0.15 c/d and $\pm$0.25 c/d for the datasets 2003
V, 2003 B and 2002 B datasets, respectively. Note that for the 2003 B
and 2002 B datasets the frequencies are not listed by decreasing
amplitude, but by correspondence with the 2003 V dataset frequencies
(see text).}
\label{tab2}
\begin{tabular}{cccc|ccc|ccc}
\hline
\noalign{\smallskip}
\multicolumn{4}{c}{\bf 2003 V dataset} & \multicolumn{3}{c}{\bf 2003 B dataset} & \multicolumn{3}{c}{\bf 2002 B dataset}\\
\hline
  & Frequency   &   Amplitude  & confidence  & Frequency   &   Amplitude  & confidence   & Frequency   &   Amplitude  & confidence \\
  & (c/d)       &    (mmag)    & (\%)        & (c/d)       &    (mmag)    & (\%)         & (c/d)       &    (mmag)    & (\%) \\
\hline
$f_1$  & 22.89     & 1.9 & 99.9  & 22.89      & 3.1 & 99.9 &  22.88    & 3.3 & 99.9  \\
$f_2$  & 34.60     & 1.5 & 99.9  & 34.82      & 2.1 & 99.9 &  34.64    & 3.3 & 99.9  \\
$f_3$  & 30.45     & 1.8 & 99.9  & 30.45      & 1.8 & 99.9 &  30.48    & 3.2 & 99.9  \\
$f_4$  & 48.23     & 1.6 & 99.9  & 48.45      & 2.0 & 99.9 &  48.23    & 1.9 & 99.9  \\
$f_5$  & 28.79     & 1.5 & 99.9  &            &     &      &  27.73    & 1.9 & 99.9  \\
$f_6$  & 23.99     & 1.3 & 99.9  &            &     &      &           &     &       \\
$f_7$  &  9.30     & 1.3 & 99.9  &            &     &      &           &     &       \\
$f_8$  & 41.11     & 1.2 & 99.9  & 42.08      & 1.9 & 99.0 &  42.27    & 2.3 & 99.9  \\
$f_9$  & 52.00     & 1.1 & 99.9  &            &     &      &           &     &       \\
\noalign{\smallskip}
\hline
\end{tabular}
\end{table*}

\section{Frequency Analysis}

\begin{figure}
   \centering
   \includegraphics[height=9cm]{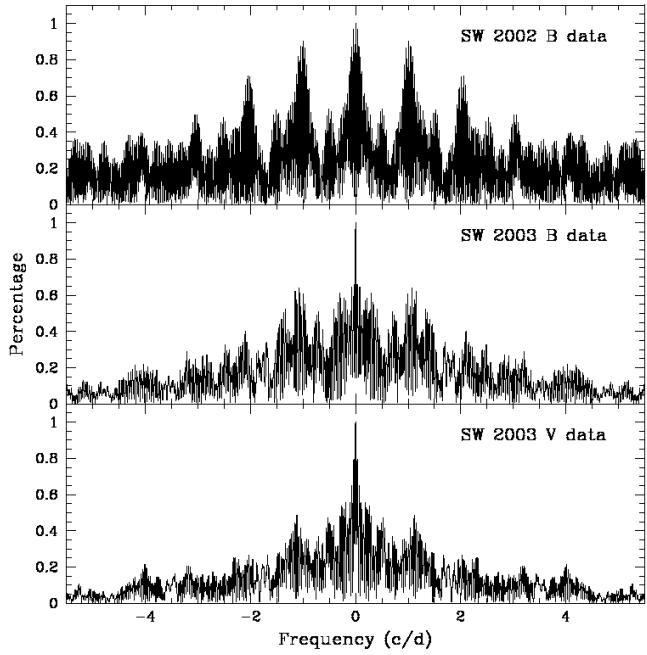}
      \caption{Spectral window in amplitude for the three data sets
analysed in this paper (see labels in the figure). Note in particular
that the 1 c/d alias is reduced to about 50\% for the 2003 V dataset.}
      \label{f3}
   \end{figure}

The frequency analysis was performed using the period04 package
(\cite{lenz}), based on the Fourier transform method. For a better
interpretation of the results, we have first calculated the spectral
window (SW) for each dataset. The result is shown in Fig.~\ref{f3}
where from top to bottom we report the SW for the three time series
identified in the previous section (see labels in the figure). The SW
was used as a diagnostic to distinguish between real and spurious
frequencies.

   \begin{figure}
   \centering \includegraphics[height=8cm]{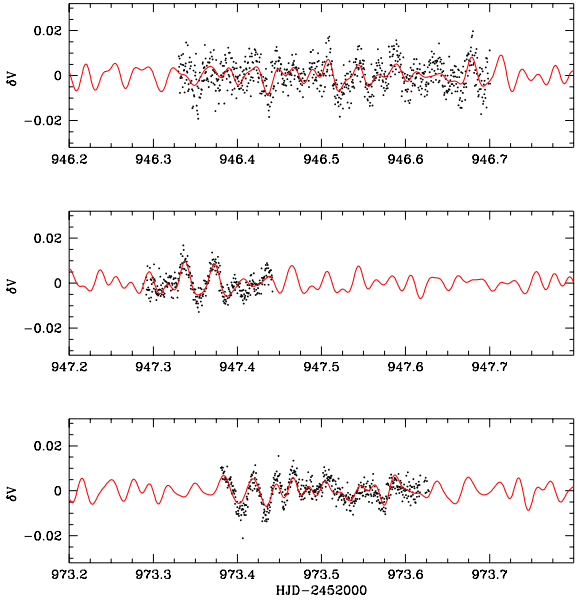}
\caption{Data collected in Loiano with the TTCP (V filter) during
Nov./Dec. 2003 (dots).  The solid line show the fit to the data with 9
frequencies obtained from the period analysis (see Sect.3).}
      \label{f2}
   \end{figure}
%

\begin{figure}
   \centering
   \includegraphics[width=9cm]{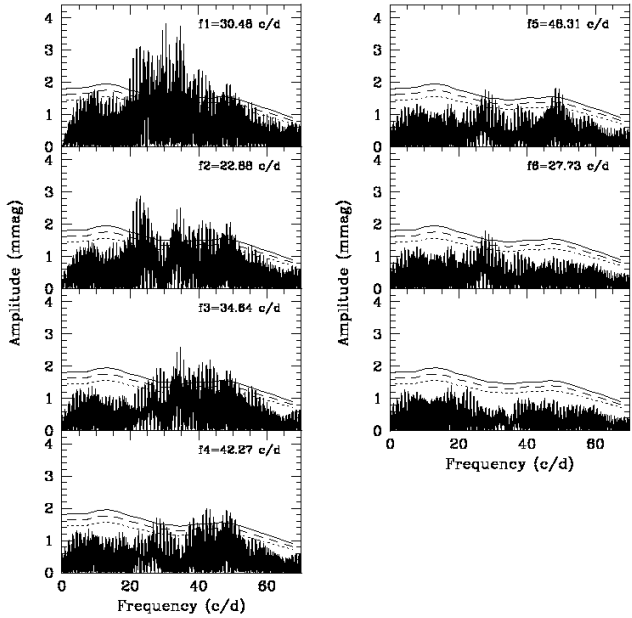}
      \caption{Frequency analysis for the 2002 B dataset. The solid,
dashed and dotted lines show respectively the 99.9\%, 99\% and 90\%
significant levels. In each panel, one peak (i.e. the labelled
frequency) is selected and removed from the time series and a new
spectrum is obtained.  The last panel shows the periodogram after the
prewithening with all the significant frequencies.}
      \label{f4}
   \end{figure}

\begin{figure}
   \centering
   \includegraphics[width=9cm]{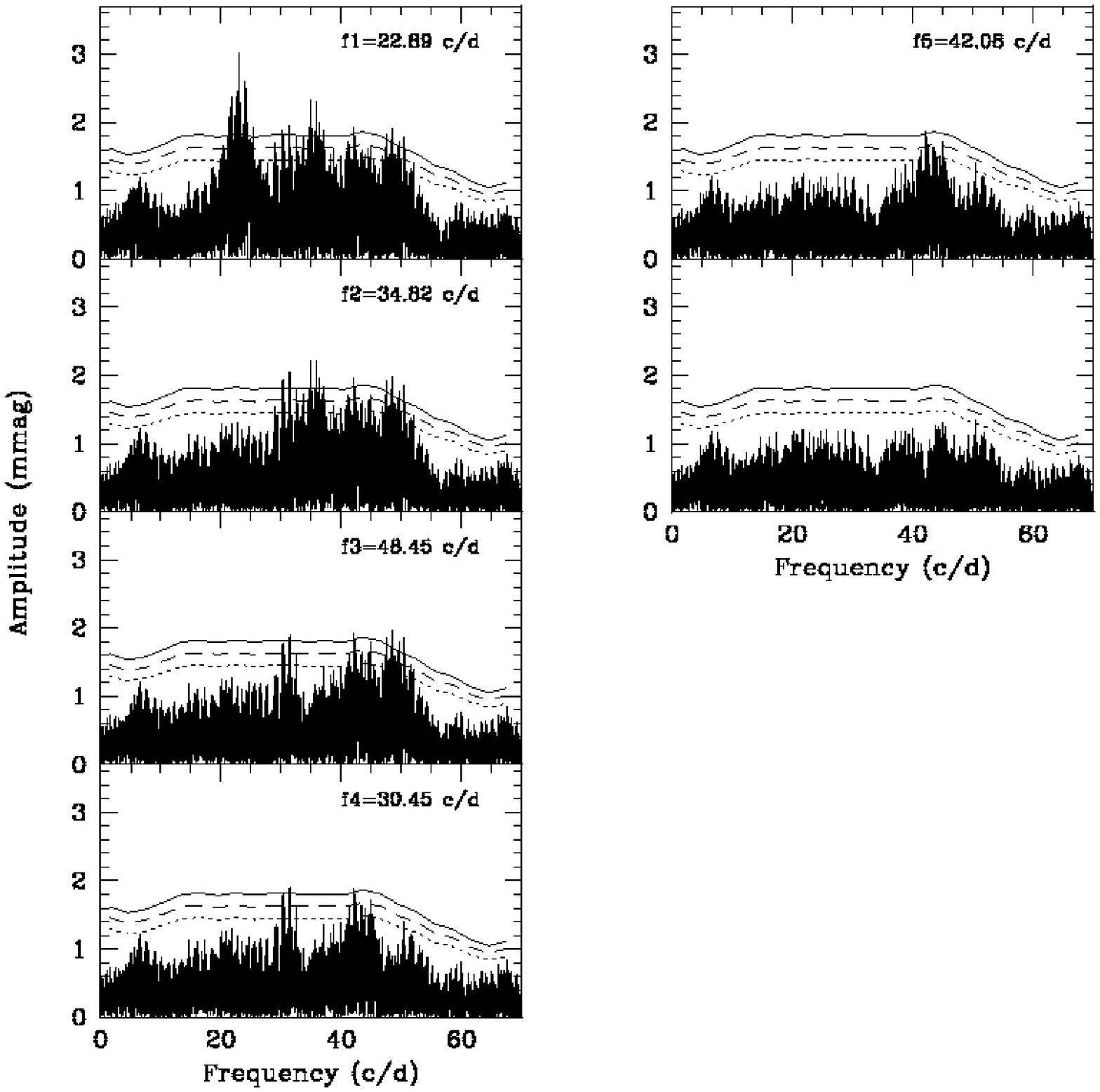}
      \caption{As in Fig.~\ref{f4} but for the 2003 B dataset.}
      \label{f5}
   \end{figure}

\begin{figure}
   \centering
   \includegraphics[width=8cm]{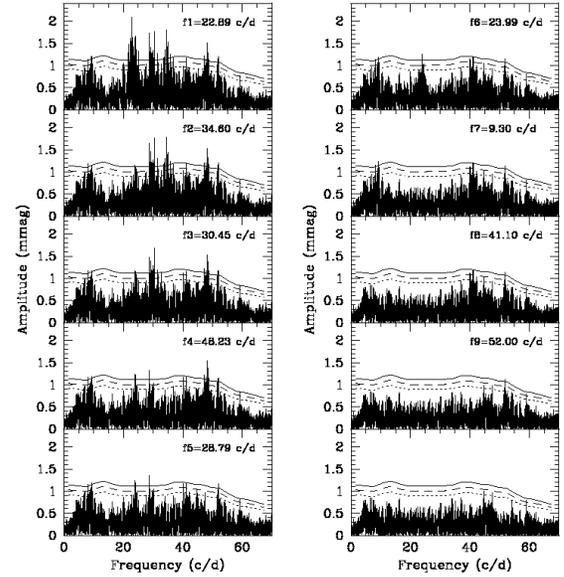}
      \caption{As in Fig.~\ref{f4} but for the 2003 V dataset.}
      \label{f6}
   \end{figure}

Each data set described in the previous Section has been analysed
separately.  Figures~\ref{f4},~\ref{f5},~\ref{f6} show the Fourier
transform for the datasets 2002 B, 2003 B and 2003 V
respectively. Here, in each panel the peak with largest amplitude
is selected and then removed, obtaining a new spectrum shown in the
following panel.  The last panel shows the periodogram after the
prewithening with all the significant frequencies. The solid, dashed
and dotted lines show the 99.9\%, 99\% and 90\% confidence levels
calculated following the widely used recipe by \cite{breger93} and
\cite{kus97}. The error on the measured frequencies (apart from the
$\pm$ 1 c/d alias) can be roughly estimated from the FWHM of the main
lobe in the spectral window (see \cite{alvarez} and references
therein). As a result we found $\Delta f$ $\sim$ 0.25 c/d, 0.15 c/d
and 0.11 c/d for the datasets 2002 B, 2003 B and 2003 V
respectively.\par The frequencies found for the three datasets are
summarized in Table~\ref{tab2}. In order to discuss in detail the
results of the frequency analysis, let us take as reference the
frequencies obtained with the 2003 V dataset which is the best
one. Then we find:

\begin{itemize}
\item
$f_1=22.89$ c/d, $f_2=34.60$ c/d, $f_3=30.45$ c/d, $f_4=48.23$ c/d are 
present, within the quoted errors, in every dataset.
\item
$f_5=28.79$ c/d is the +1 c/d alias of the $f=27.73$ c/d frequency found  
in the 2002 B dataset. Due to the better duty cycle of the V data we 
shall retain $f_5=28.79$ c/d for comparison with the models. 
\item
$f_8=41.11$ c/d is probably the -1 c/d alias of 
$f=42.08$ c/d in the 2003 B dataset and 
$f=42.27$ c/d in the 200V B dataset. As before,
 we shall use the frequency obtained in the V filter. 
\item
$f_6$=23.99 c/d, $f_7$=9.30 c/d and $f_9=52.00$ c/d are found only in
the V filter probably because of the better S/N of this dataset. 
\end{itemize}

In summary, the multisite campaign allowed us to identify up to 9
significant frequencies for IP Per. The fit to the observed data
points obtained with period04 including the nine is shown in
Fig.\ref{f2}.  Five of these are present in each dataset. In
particular we confirm all the six frequencies (within the errors and
taking into account the 1 c/d alias) found during the single site CCD
study (i.e. the 2002 B dataset) of IP Per.

%

%

%

%

\section{Comparison with theoretical models}

\begin{figure}
   \centering
   \includegraphics[width=8cm]{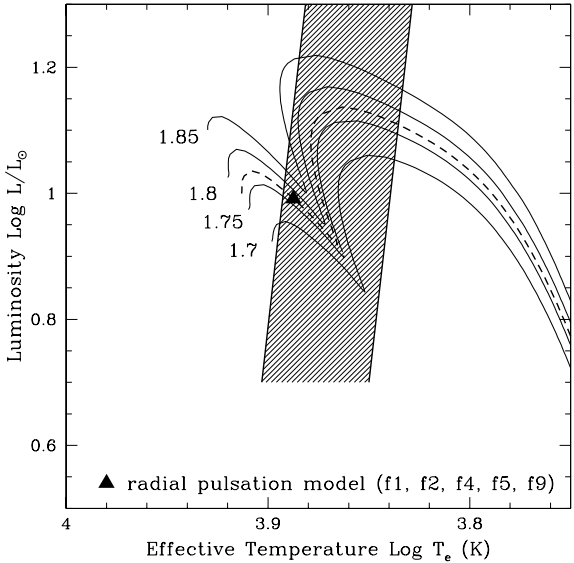}
      \caption{Position in the HR diagram of the best-fit radial
pulsation model.  The shaded region is the predicted instability strip
by \cite{marconi} with solar chemical composition (Z=0.02, Y=0.28).
Solid lines are the PMS evolutionary tracks computed for the labelled
stellar masses with the FRANEC stellar evolution code (\cite{chieffi},
\cite{castellani}). The dashed lines represent the 1.77$M_{\odot}$ PMS
track with solar chemical composition.  }
      \label{f7}
   \end{figure}

The results presented in the previous Section can be used to constrain
the intrinsic stellar properties of IP Per and in particular its mass
and position in the HR diagram, through comparison with stellar
pulsation models.  Using a linear non-adiabatic pulsation code (see
Marconi \& Palla 1998, \cite{marconi2004a} for details) we could not
reproduce all the observed frequencies. In fact, we can recover at
most 5 of the 9 observed frequencies for $M=1.77\pm0.01 M_{\odot}$,
$\log{L/L_{\odot}}=0.992\pm0.003$, $\log T_{\rm
eff}=3.887\pm0.002$. This solution corresponds to a radial pulsation
model which simultaneously oscillates in the first ($f_1$), second
($f_5$), third ($f_2$), fifth ($f_4$) and sixth ($f_9$) overtones. Its
position in the HR diagram is shown in Fig.~\ref{f7} together with the
predicted instability strip by \cite{marconi} and the PMS evolutionary
tracks computed for the labelled stellar masses with the FRANEC
stellar evolution code (\cite{chieffi}, \cite{castellani}).  The $1.77
M_{\odot}$ PMS track is represented by the dotted line.

In order to reproduce the observed frequencies that were not found
from the radial analysis and, at the same time to investigate the
possibility, strongly supported by empirical evidence on other PMS
$\delta$ Scuti stars (see e.g. \cite{balona}), that non-radial modes
are also present in IP Per, we have attempted an asteroseismological
interpretation of the data using the Aarhus adiabatic non-radial
pulsation code (http://astro.phys.au.dk/$\sim$jcd/adipack.n/).  A
preliminary application of this code to the PMS evolutionary structure
corresponding to the best-fit radial pulsating model seems to suggest
that $f_3$, $f_6$ and $f_8$ are associated to non-radial modes with
$l=2$. As for $f_7$, neither the radial nor the non-radial analyses
are able to match the observed value for the selected stellar
parameters.

%

\subsection{Comparison with post-MS pulsation models}

Both spectroscopic measurements and PMS model predictions seem to
suggest that the position of IP Per in the HR diagram is near the MS,
in a region where PMS and post-MS evolutionary tracks, at fixed
stellar mass, are known to intersect and remain quite close to each
other. In order to investigate the effect of the assumed evolutionary
status on the predicted frequencies, we have taken into account the
post-MS evolution of stellar models with masses ranging from 1.7 to
1.8 $M_{\odot}$, computed with the FRANEC code. As a result, we
find that the post-MS evolutionary model located at the same
luminosity and effective temperature of our PMS best fit model (see
Fig.~\ref{f7}), has a stellar mass of 1.73 $M_{\odot}$, i.e. only slightly
lower than the PMS counterpart (1.77 $M_{\odot}$). As a consequence
the structure of this post-MS model produces periods slightly longer
than the best fit PMS solution, but differences are of the order of
minutes for both radial and non-radial $p$ modes. The 1.73 $M_{\odot}$
post-MS model, reproducing the same number of frequencies as the PMS
one, would require a slightly different luminosity and effective
temperature, namely $\log{L/L_{\odot}}=0.988$, $\log{T_{\rm
eff}}=3.888$. The values of the predicted $p$ mode periodicities are
in this case very similar to the PMS ones. However, we know that the
small differences in the non-radial frequencies can produce evident
changes in the large and small frequency separations, in particular
for $l=2$, as extensively discussed by \cite{suran}. This occurrence
would in principle allow to verify the PMS nature of IP Per if very
accurate and long time based observations (e.g. from space) were
available.

\subsection{Effect of the metal abundance}

All the above evolutionary and pulsational analysis is based on models
with solar chemical composition, namely Z=0.02, Y=0.28.  If the
evidence pointed out by \cite{miro} that IP Per has significantly
lower metal abundance ($Z \simeq 0.008$) is confirmed, the above
theoretical interpretation would have to be modified, as discussed
in the following.

\subsubsection{The instability strip}

\begin{figure}
   \centering
   \includegraphics[width=8cm]{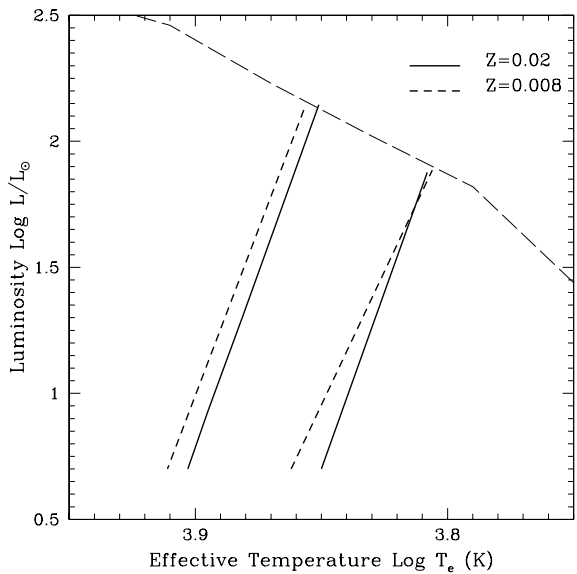}
      \caption{Position of the instability strip for the three radial
modes at Z=0.008 (dashed lines) compared with the one at solar
metallicity (solid lines). The long dashed line represent the 
birthline by \cite{palla93}.}
      \label{stripZ}
   \end{figure}

  We computed new nonlinear radial pulsation
models, with the same code and the same numerical and physical
assumptions as in Marconi \& Palla (1998), but with Z=0.008,
Y=0.25. The resulting instability strip for the first three radial
modes is reported in Fig.\ref{stripZ} (dashed lines). As shown in this
figure, where the instability strip for Z=0.02 is also plotted for
comparison, the metallicity effect on the theoretical boundaries is
rather small (200 K at most for each luminosity level) and decreases
toward the higher luminosities, with a global shift of the instability
region toward higher effective temperatures, as Z decreases from Z=0.02
to Z=0.008.  

\subsubsection{Pulsational frequencies}

\begin{figure}
   \centering
   \includegraphics[width=8cm]{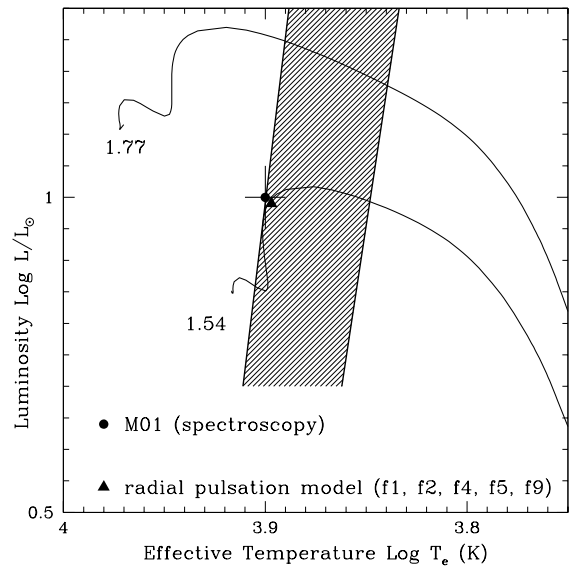}
      \caption{Best fit radial pulsation model for Z=0.008 in the HR
diagram (filled triangle) as compared with the spectroscopic
measurement by \cite{miro}. The corresponding PMS evolutionary track
and instability strip are overplotted. The track for 1.77 $M_{\odot}$,
Z=0.008 is shown for comparison with Fig. \ref{f7}.  }
      \label{bestfitZ}
   \end{figure}

We also computed new PMS
evolutionary models at Z=0.008, Y=0.25 with the FRANEC code and the
same assumptions used for solar chemical composition. Then we computed
linear nonadiabatic radial pulsation models\footnote{We did not
compute nonradial models because our version of the Aarhus nonradial
code assumes solar metallicity. The extension of nonradial
computations to other chemical compositions will be addressed in a
future work.}  along the Z=0.008 PMS tracks, in order to match the
observed frequencies of IP Per. The resulting best fit radial model
for Z=0.008 Y=0.25, again pulsating in five radial modes with $f_1$,
$f_2$, $f_4$, $f_5$ and $f_9$, corresponds to a PMS structure with
$M=1.54\pm0.01 M_{\odot}$, $\log{L/L_{\odot}}=0.991\pm0.003$, $\log
T_{\rm eff}=3.896\pm0.002$.  Its position in the HR diagram is shown
in Fig.~\ref{bestfitZ} together with the predicted instability strip
for Z=0.008 and the corresponding PMS evolutionary track.  Inspection
of this figure clearly indicates that the predicted position in the HR
diagram of IP Per, assuming Z=0.008, is in good agreement with the
empirical spectroscopic determination by \cite{miro} (filled circle).\par

\medskip

However, the observationally derived metal deficit reflects only the
current state of the atmosphere and may be due to recent effects, such
as interaction with the circumstellar medium (see \cite{Gray}).
In this case it would represent only a surface effect. As the
PMS structure of our best fit model is characterized by a radiative
envelope, such a contamination should not have any effect on the
pulsation properties which involve deeper layers across the Hydrogen
and Helium ionization regions.On the other hand the different
abundances in the surface layer would affect the position in the HR
diagram due to the opacity variation. A detailed modeling of this case
is beyond the purposes of the present paper. At the same time a more
accurate determination of the metal poor nature of IP Per 
would be important.

%

%

\section{Conclusions}

A total of about 190 hours of observations obtained during 38 nights
at 9 different telescopes on the PMS $\delta $ Scuti star IP Per have
been presented. The Fourier analysis of this data set confirms the
multiperiodic nature of this pulsator: we have identified nine
frequencies of pulsation which are significant on the basis of the
\cite{breger93} and \cite{kus97} criteria.

Comparison with the predictions of linear non-adiabatic radial
pulsation models allows us to recover only five out of the nine observed
frequencies for a 1.77 $M_{\odot}$ model.  Non-radial pulsation is
also present in this star. A preliminary interpretation of the
observed frequencies through the Aarhus non-radial code, applied to
the evolutionary structure of the 1.77 $M_{\odot}$ model reproducing
$f_1$, $f_2$, $f_4$, $f_5$ and $f_9$ with radial modes, seems to
indicate that $f_3$, $f_6$ and $f_8$ are associated with non-radial
modes with $l=2$. Specific post-MS evolutionary and pulsational models
were computed in order to investigate the dependence of radial and
non-radial output frequencies on the assumed evolutionary status. The
resulting post-MS solution has similar stellar parameters and $p$ mode
frequencies.\par

\medskip
Finally the possible effect of the metal poor nature of IP Per
detected by \cite{miro} on both pulsation and evolutionary properties
is discussed.  We find that if the metallicity of IP Per is as low as
Z=0.008 the best fit radial model has a significantly lower mass than
the case at solar chemical composition but the pulsation
characteristics are similar.Also the estimated position of IP Per in
the HR diagram appears to be in good agreement with the independent
determination by \cite{miro}.  
Whether the low metallicity is a property only of the surface layers
or represents a systematic deficit throughout the interior, as we have
assumed in our modeling, should be clarified before final conclusions
on the stellar parameters of IP Per can be reached.

\begin{acknowledgements}
We wish to thank our referee, Dr. Miroshnichenko for his valuable
suggestions which  helped  improving the paper.
This work made use of CDS database in Strasbourg. It is a pleasure to thank 
J. Christensen-Dalsgaard for useful comments and suggestions 
on the use of the Aarhus adiabatic non-radial pulsation code. We also
thank S. Degl'Innocenti and P.G. Prada Moroni for their help with 
the FRANEC code. We are indebted to S. Leccia for useful discussions. 
V.R. wishes to thank the personnel of the Loiano Observatory for their 
help with the observations. T.D.O. acknowledges support from NSF 
grant AST0206115.

\end{acknowledgements}

\appendix

\section{Details on the instrumentation used}

In this appendix we describe in some detail the 
combinations telescope/instrument used during the 2003 multisite 
campaign and summarized in Tab.~\ref{tab1}. 
An indication on data reduction procedures is reported too. 

\begin{enumerate}

\item
The Loiano (2003) observations were obtained with the 1.5m 
telescope equipped with the Toppo Three Channel Photometer 
(TTCP, see http://www.na.astro.it/\%7Esilvotti/TTCP.html for details). 
This instrument allows the simultaneous observation of 
the variable and comparison stars, as well as of the sky. Therefore 
it is possible to obtain high precision photometry 
with fast sampling. The data reduction is also straitforward. 

\item 
The BAO measurements were obtained with the 0.85m telescope placed 
at the Beijing Astronomical Observatory. The telescope is equipped with a 
Three Channel Photometer (TCP) which is very similar to the TTCP quoted above.
More details on BAO instruments can be found at www.bao.ac.cn. 

\item
The OSN measurements were obtained with the 0.90 m telescope 
located at the Observatorio de Sierra Nevada in Granada, Spain. 
The telescope is equipped with a simultaneous four-channel 
photometer ($uvby$ Str\"omgren photoelectric photometer). More details 
on the instrument can be found at http://www.osn.iaa.es/fotometro\_eng.html. 

\item
The SPM measurements were carried out with the 1.5 m telescope 
at the San Pedro Martir Observatory, Mexico. The 1.5 m telescope was equipped 
with a simultaneous $uvby$ Str\"omgren photoelectric photometer, which is the 
twin of the OSN instrument mentioned above.
The SPM instrument (named ``Danish'') is described at 
http://haro.astrossp.unam.mx/Instruments/.

\item
The Serra la Nave data were obtained with the 0.91cm telescope of the 
Catania Astrophysical Observatory, Italy. The telescope is equipped 
with a Single Channel Photometer (SCP here after) which is described at 
http://w3c.ct.astro.it/sln/strumenti.html.  

\item
The APT measurements were obtained with the T6 0.75 m 
Vienna Automatic Photoelectric Telescope (APT), situated at 
Washington Camp in Arizona, USA (\cite{strass};\cite{breger99}).   
Also this telescope is equipped with a SCP.

\item
The SARA measurements were collected by using the 0.9 m automated telescope 
at Kitt Peak in Arizona (USA). 
Images were collected using an Apogee AP7p camera with a
back-illuminated SITe SIA 502AB 512x512 pixel CCD.  The pixels 
are 24 microns square, corresponding to 0.73" at the telescope
focal plane scale.  
Sky flats, dark and bias exposures were taken every night.  All data
was calibrated and reduced using standard IRAF routines.

\item
The OGS observations were collected with the 1.0m OGS telescope placed at 
the Teide Observatory (Tenerife, Canary Islands, Spain). The telescope 
was equipped with a 1024x1024 pixel CCD. 
The pixel scale was 0.32$^{''}$ for a total field of view 
of 5.5$^{'}$x5.5$^{'}$. 
Typical integration  times were 10-15 sec, depending upon sky conditions.
Sky flats, dark and bias exposures were taken every night.  All data
was calibrated and reduced using standard IRAF routines.

\item
The SOAO data were obtained with the 0.61m telescope placed at the
Sobaeksan Optical Astronomy Observatory (SOAO) in Korea.  The
telescope was equipped with a SITe 2K CCD camera.  The field of view
of a CCD image is about $20.5 \times 20.5$ arcmin$^{2}$ on the f/13.5
Cassegrain focus of the telescope. Data reduction has been performed
using the standard IRAF/CCDRED package.  Instrumental magnitudes were
obtained via the simple aperture photometric routine in the
IRAF/APPHOT package.

\end{enumerate}

\end{document}